\documentstyle[12pt,aasms4,flushrt]{article}

\begin{document}

\title{Are He and N abundances in Type I PNe as high as empirically 
derived?}
\author{Ruth Gruenwald and Sueli M. Viegas}
\author{Instituto Astron\^omico e Geofisico - USP}
\author{Av. Miguel Stefano, 4200}   
\author{04301-904 S\~ao Paulo, Brazil} 

\begin{abstract}
Type I planetary nebulae (PNe) are defined as those  with high He and N 
abundances (Peimbert $\&$ Torres-Peimbert 1983). These objects present 
in general bipolar geometries and 
have high stellar temperatures (Corradi $\&$ Schwarz 1995, Torres-Peimbert
$\&$ Peimbert 1997). In this paper we analyse the 
empirical methods for abundance determination in order to check if the 
He and N overabundances in Type I PNe are a consequence of a geometrical 
effect, due to the bipolarity, or the
ionization stratification, due to the stellar temperature. For this, 
we obtain simulated spherically symmetrical as well as bipolar nebulae, using 
a 3D photoionization code. From the 
projected emission line intensities for: a) the whole nebula; b) for a slit 
crossing the nebula;  as well as c) for different positions in the nebula, 
we applied the formulae used in the literature to obtain empirical 
abundances.  These  empirical abundances are then compared with 
the adopted ones. We show that empirical 
abundances depend on the particular line of sight covered by the observation
and can simulate an 
overabundance and/or the presence of abundance gradients of He and N in 
planetary nebulae with high stellar temperature. The geometrical effects are 
also discussed. Systematic errors in abundance determinations by empirical 
methods are higher for the N/H ratio than for N/O. Thus, it seems better to 
use the N/O value when discussing N rich objects.
\end{abstract}

\section{Introduction}

The knowledge of chemical abundances in Planetary Nebulae (PNe) is 
important since they can be used as constraints for stellar 
evolutionary models, as well as for studies of the chemical evolution 
of galaxies. 
In fact, PNe abundances reflect the products of stellar evolution, as well   
as the enrichment of the interstellar medium. 
Chemical abundances in PNe are usually 
obtained using empirical methods (as for example in Peimbert $\&$ 
Torres-Peimbert 1987, Kingsburgh $\&$ Barlow 1994). 
Empirical methods are commonly used since they can easily be applied to 
a great number of objects, from a small number of bright emission lines.
Photoionization models also provide the gas chemical composition 
when applied to specific nebulae (for example, Harrington 
et al. 1982; Clegg et al. 1987; Gruenwald, Viegas $\&$ Brogui\`ere 1997). 

Following their abundance, PN are classified in types
(Peimbert 1978; Peimbert $\&$ Torres-Peimbert 1983; Fa\'undez-Abans $\&$ 
Maciel 1987). 
Type I PNe are those with  
high He and N abundances, e.g. He/H $\geq$  0.125 and log(N/O) $>$ -0.30
(Peimbert $\&$ Torres-Peimbert 1983).
In a study of morphological and physical properties of PNe, 
Corradi $\&$ Schwarz (1995) state that 
bipolar nebulae have the hottest stars among PNe, and, except for two 
objects, all bipolar PNe for which chemical data are available are Type I. 
Another common characteristic of 
bipolar nebulae is that the [NII] line intensity is stronger than H$\alpha$ 
all over the nebula, apart from the very central region 
(Corradi $\&$ Schwarz 1993b).
This characteristic is usually 
explained by a N overabundance or by shock heating (Corradi $\&$ 
Schwarz 1993a; Corradi $\&$ Schwarz 1993c).

Predictions of stellar evolutionary models and the high values obtained 
for He and N abundances in Type I/bipolar planetaries have suggested 
that their central stars are more massive than other PN nuclei
(Peimbert $\&$ Serrano 1980; Calvet $\&$ Peimbert 1983; Torres-Peimbert 1984; 
Corradi $\&$ Schwarz 1995). However, new evolutionary models 
fail to explain the high abundances of He and N in Type I PNe (Marigo, Bressan
$\&$ Chiosi 1996), although models including the ``hot botton burning'' 
effect and accounting for the chemical evolution on the AGB may lead to 
higher N/O abundances (van den Hoek $\&$ Groenewegen 1997). 

From the characteristics of Type I/bipolar planetary nebulae,
it can be noted that objects with the highest He and N empirical abundances 
are those presenting a bipolar geometry and having the highest stellar 
temperatures. In fact, objects with higher stellar 
temperatures (T$_*$) tend to present higher He and N empirical abundances
(see Fig. 1). 
In order to homogenize the data in the figure, only objects with 
abundances obtained from 
line intensities integrated in a slit across the nebulae are plotted.
As it will be shown in the following sections, empirical abundance 
determinations depend on the line of sight covered by the observation.
Empirical abundances for the objects plotted in Fig. 1 are from 
de Freitas Pacheco et al. (1992), Kingsburgh $\&$ Barlow (1994), and Costa 
et al. (1996). The stellar temperature
corresponds to the Zanstra temperature of HeII.
Values for T$_*$ are taken from the literature (Kaler 1983; Preite-Martinez 
$\&$ Pottasch 1983; Pottasch 1984; Shaw $\&$ Kaler 1985; Gathier $\&$ 
Pottasch 1988; Gathier $\&$ Pottasch 1989; Gleizes, Acker, $\&$ Stenholm 1989;
Kaler $\&$ Jacoby 1989; Shaw $\&$ Kaler 1989; Kaler, Shaw, $\&$ Kwitter 1990; 
Stanghellini, Corradi, $\&$ Schwarz 1993).

In this paper we analyze the effect of the geometry and of the stellar 
temperature on the empirical abundance determinations, in order to verify 
if the high abundances of He and N obtained for Type I PNe are real. 
In fact, several studies show that the stellar temperature and/or 
the geometry can affect the abundances obtained from empirical methods.
As discussed by Peimbert and collaborators (Peimbert 1967; 
Peimbert $\&$ Costero 1969; Peimbert, Luridiana, $\&$ Torres-Peimbert 1995), 
the gas temperature obtained from observed forbidden line ratios 
and used in abundance determinations must be corrected by 
temperature fluctuations inside the nebulae. 
In fact, for typical conditions of PNe, temperature 
fluctuations can be important 
when the ionizing central star is hotter than 10$^5$ K (Gruenwald $\&$ 
Viegas 1995).
Furthermore, a study on spherically symmetrical homogeneous HII regions 
show that temperature fluctuations  obtained from projected data vary
with the line of sight to the nebula; the variation 
depending on the characteristics of the star and the nebula (Gruenwald 
$\&$ Viegas 1992). 
With a three-dimensional (3D)
self-consistent photoioinization code, 
a model for the bipolar PN IC 4406 has been obtained, showing that its 
abundance does not fulfill the criteria for Type I, as this nebula is usually 
classified (Gruenwald et al. 1997). Such an analysis shows that specific 
and more realistic models which take into 
account the geometry of the object can lead to abundance 
results that differ from 
those obtained with empirical methods.

In order to check for the effects of geometry and stellar temperature on the 
empirical method to obtain PNe chemical abundances, we apply 
our self-consistent 3D photoionization code to 
planetary nebulae. For a wide range of nebular and  stellar 
characteristics, we obtain the physical conditions in each point of the 
nebula, and calculate the resulting line intensity ratios at
different lines of sight to the nebula. From the calculated line intensity 
ratios we determine the abundances applying the empirical methods used 
in the literature. These abundances are then compared to those assumed 
in the models. The models are described in \S 2. The results are 
presented in \S 3 and the conclusions are outlined in \S 4. 

\section{Empirical Abundances from Simulated Nebulae}

The 3D code has been described in a previous paper (Gruenwald et al. 1997).
As in classical 
uni-dimensional photoionization codes,
the input parameters are the stellar temperature and luminosity, 
as well as the nebular elemental abundances and  density distribution. 
Typical parameters for planetary nebulae are assumed: 
stellar temperatures (T$_*$) ranging
from 2$\times$10$^4$ to 3$\times$10$^5$ K, and stellar luminosities (L$_*$) 
from 3$\times$10$^2$ L$_\odot$ to 2$\times$10$^4$ L$_\odot$.
The models are obtained for two sets of chemical abundances (uniform 
throughout the nebula): 
solar abundances (Grevesse $\&$ Anders 1989),  Z$_{\odot}$, 
and an average 
abundance for planetary nebulae (Kingsburgh and 
Barlow 1994), Z$_{typ}$, which includes 
He, C, N, O, Ne, S, and Ar; for the remaining elements 
(Mg, Si, Cl, and Fe), we assume a value equal to a hundredth solar 
in order to take into account their presence in grains (Stasinska $\&$ 
Tylenda 1986).
For spherically symmetrical nebulae, a constant density, n$_H$, ranging 
from 10$^2$ to 10$^6$ cm$^{-3}$ is assumed. 
In order to analyze PNe with bipolar apparence, 
an equatorial torus of denser gas close to the star is included.
The presence of a dense equatorial region can explain the observed geometry 
of bipolar PNe (Calvet $\&$ Peimbert 1983; Icke, Preston, $\&$ Balick 1989;
Sahai et al. 1991; Corradi $\&$ Schwarz 1995; 
Gruenwald et al. 1997).  
The torus has the following characteristics:  
i) external radius, r$_{ext}$, equals to R/10, 
where R is the maximum dimension of the ionized region; 
ii) internal radius, r$_{int}$,
equal to r$_{ext}$/10; iii) height, h$_{disk}$, equal to r$_{int}$; 
iv) n$_{torus}$= 10 n$_H$. 
The maximum dimension R corresponds to the distance from the star where
the fractional ionization of H, H$^+$/H, equals to 5$\%$, along the torus axis.
The torus 
is centered at the central star. These  values for the torus dimensions 
and density are chosen just to analyze the 
geometry effect on the abundance determinations, and are not intended to  
represent all the bipolar nebulae.

The code provides the emissivity of the lines in each point of the nebula.
Line intensity ratios can then be obtained: 
a) for different positions in the projected nebula;
b) for the whole nebula; 
c) for a given slit across the nebula.  

We assume that the calculated line intensity ratios simulate 
observed ratios. Therefore, 
chemical abundances can be obtained by empirical methods using 
these line ratios.
Comparing the calculated abundances with those adopted 
for the simulated nebula,
the possible effects of geometry and stellar temperature
on the empirical abundance calculation  can be established.
The calculations are performed for 
different sets of emission line intensity ratios 
corresponding to cases (a), (b), and (c), listed above.

In order to obtain empirical elemental abundances, the ionic fractional 
abundance must be computed. For ions producing 
detected emission lines, the ion abundance relative to H$^+$ is obtained 
from the observed line intensity relative to H$\beta$ and the corresponding 
line emissivity 
ratio. Line emissivities depend on the gas density and temperature.
The electronic density is obtained from the 
[SII] $\lambda$6718/$\lambda$6733 line 
intensity ratio, since this ratio depends weakly on the gas temperature 
(Osterbrock, 1989).
Knowing the density, the gas temperature for the high ionization 
region (T$_{[OIII]}$) can be obtained from the 
[OIII] $\lambda$$\lambda$4959+5007]/$\lambda$4363 line ratio, 
while for the low 
ionization region (T$_{[NII]}$) from the 
[NII] $\lambda$$\lambda$6548+84]/$\lambda$5755 line ratio.
Line emissivities, 
electronic density, and temperature are calculated assuming a five-level atom.

A correction must be done for ions which are present in the gas and do 
not produce detectable emission lines. 
Such a correction can be done using the corresponding ``ionization correction 
factor'' ({\it icf}) for each element. We apply the 
{\it icf} from Peimbert $\&$ Torres-Peimbert (1987), as generally
used in the literature. Empirical 
abundances for He, O, and N are then  obtained from the following equations:

\begin{equation}
\frac{He}{H} = \frac{He^{+} + He^{++}}{H^{+}} 
\end{equation}

\begin{equation}
\frac{O}{H} = icf(O) \Biggl(\frac{O^{+} + O^{++}}{H^{+}}\Biggr)
\end{equation}

\begin{equation}
\frac{N}{H} = \Biggl(\frac{N^{+}}{O^{+}}\Biggr) \Biggl(\frac{O}{H}\Biggr) = 
\Biggl(\frac{N^{+}}{H^{+}}\Biggr)
\Biggl(1 + \frac{O^{++}}{O^{+}}\Biggr) \Biggl(1+ \frac{He^{++}}{He^{+}}\Biggr)
\end{equation}

\par\noindent
where \begin{equation} icf(O) = \frac{He^{+}+He^{++}}{He^{+}}.
\end{equation}

The ionization correction factor for O, $icf$(O), accounts for ionization 
stages higher 
than O$^{++}$. In equation (3), the last two terms 
correspond to the correction due, respectively, to N$^{++}$ and to higher 
stages of N. 

The abundance of He$^+$ and He$^{++}$, relative to H$^+$, are determined 
respectively from the HeI $\lambda$5876
and HeII $\lambda$4686 line intensities relative to H$\beta$. The ratio of 
line emissivities for these lines were calculated following Barker (1978), 
using the values obtained for the gas density and temperature. 
A correction due to collisional excitation of He$^+$ is taken into account 
following Peimbert $\&$ Torres-Peimbert (1987). 
The O$^+$, O$^{++}$, and N$^+$ abundances. relative to H$^+$, 
are obtained respectively from   
[OII] $\lambda$$\lambda$3727+29/H$\beta$, 
[OIII] $\lambda$$\lambda$4959+5007/H$\beta$, and
[NII] $\lambda$$\lambda$6548+84/H$\beta$ line ratios.

\section{Results}

In the following, the emission line ratios from simulated nebulae and 
the corresponding empirical chemical abundances are discussed for
cases (a), (b), and (c) listed above.

\subsection{Spherically symmetrical nebulae}

The behavior of the [NII] $\lambda\lambda$ 6548+84/H$\alpha$ 
line intensity ratio with the projected 
distance to the center of the nebula is shown in Fig. 2a,
while Fig. 2b gives the corresponding N/H abundance ratio. 
The simulations are performed 
with uniformly distributed density equal to 10$^2$ cm$^{-3}$, 
Z = Z$_{typ}$, and stellar luminosity L$_*$ = 3$\times$10$^3$ L$_{\odot}$.
The projected distance to the center of the nebula is given 
in units of the radius of the ionized region.
The curves are labeled by the stellar 
temperature, in units of 10$^3$ K.
The [NII]/H$\alpha$ line ratio depends on 
the position of the line of sight to the nebula since different lines of 
sight cross different proportions of [NII] and H$\alpha$ emitting 
regions (Fig. 2a).
The behavior of the [NII]/H$\alpha$ line ratio for 
different lines of sight depends on the stellar temperature. 
Higher stellar temperatures lead to larger recombination regions, 
increasing the H$^+$ emitting volume relative to N$^+$ emitting volume.  
Therefore, there is a shift of the peak of [NII]/H$\alpha$ towards 
smaller r$_{proj}$/R with increasing T$_*$, as shown in Fig. 2a. 
The maximum value for [NII]/H$\alpha$ depends on the value 
of the nitrogen abundance used in the models. However, as seen in the figure, 
the measured line ratio can substantially vary even for a nebula with 
a homogeneous abundance. 

Details on the [NII]/H$\alpha$ distribution on the projected nebula depend
also on the stellar luminosity, on the nebular density, as well as on the 
chemical abundance. 
For increasing values of the nebular density, the region where 
[NII]/H$\alpha$ is higher than one are relatively narrower with the peak 
shifted to higher r$_{proj}$. The peak is also shifted to higher r$_{proj}$
for increasing L$_*$. Lower values for the iron abundance increases the 
[NII]/H$\alpha$ ratio since Fe is an important cooler in the region where 
[NII] lines are produced. This effect is more significant for lines 
of sight far from the central regions.

Following Corradi $\&$ Schwarz (1993b), 
[NII] stronger than H$\alpha$ appears to be a common characteristic 
of bipolar nebulae.
Our results show that even for a nebula with solar N/H abundance 
(which is lower than the average N/H for PNe),  
the measured [NII]/H$\alpha$ line ratio can be higher than one. 
Indeed, a simulated nebulae with solar abundances  
and high T$_*$ (typical of Type I planetary nebulae) present 
[NII]/H$\alpha$  higher than one all over the nebula.
For NGC 2440, which is a PN with a 
high value for T$_*$, the [NII] 
line is brighter than H$\alpha$, and the line ratio increases from 3 
to 30, from the central to the outer zones (Icke et al. 1989).
In fact, using characteristic parameters for this object 
(T$_*$ = 2$\times$10$^5$ K, L$_*$ = 5$\times$10$^3$ L$_{\odot}$, and
n$_H$ = 6$\times$10$^3$ cm$^{-3}$), and for Z = Z$_{typ}$, we verify that 
the [NII]/H$\alpha$ ratio 
can reach values of the order of 26. Notice that these values are given just 
as an illustration, since 
these results correspond to a spherical symmetrical nebula, while  NGC 2440 is 
clearly irregular (see, for example, Schwarz, Corradi $\&$ Melnick 1992).  

The variation of the empirical N abundance 
in different lines of sight to a simulated nebula, is shown in Fig. 2b.
The empirical values of N/H are plotted relative to the adopted 
abundance ratio.
The empirical value obtained for the abundance depends on the detailed 
ionization structure of the nebula. 
Notice that for PNe whose central stellar temperature is 
high (T$_*$ $>$ 10$^5$ K), the obtained values for different 
lines of sight can significantly vary, and can be higher 
than the adopted one in a extended region, mimicking an unreal overabundance, 
as well as an abundance gradient.
Therefore, studies of abundance gradients of nitrogen across a 
nebula applying empirical methods to the observed line intensities 
(Chu et al. 1991; Guerrero, Stanghellini, $\&$ Manchado 1995; 
Guerrero, Manchado, $\&$ Serra-Ricart 1996; 
Guerrero 1997; Corradi et al. 1997) can be misleading. 
For T$_*$ $\leq$ 10$^5$ K, the resulting abundances
are not very different from the adopted one, except in a narrow region 
near the borders. 
For such stellar temperatures, the volume of the H$^o$ zone is 
larger than the N$^o$ one. Lines of sight with increasing distance 
from the center include a higher fraction of low ionized regions. 
Therefore, empirical methods which do not account for neither N$^o$ nor H$^o$ 
overestimates the N abundance and must not be applied in this case.
For models with different n$_H$ and/or L$_*$, the discrepancies from the 
adopted abundance are somewhat different. However, the major effect comes 
from the stellar temperature.  

The variation of the [NII]/H$\alpha$ with T$_*$ is shown in Fig. 3a 
for the whole nebula and for a long and narrow slit across its central part.
The two upper curves correspond to simulated nebulae with Z$_{typ}$, 
while the lower ones to Z$_{\odot}$.
The long-dashed and solid lines represent line ratios for the entire nebula, 
while the 
dot-dashed and short-dashed curves correspond to line intensities 
integrated in a slit.
The nebular density and the stellar luminosity are the same as in 
Fig. 2.
As expected, 
[NII]/H$\alpha$ values are lower for a slit than for the whole nebula. 
In fact,  
in the latter case there is a higher contribution of low ionized regions, 
increasing the [NII] contribution relative to H$\alpha$, for a given T$_*$.

Results for the empirical N/H relative to the adopted abundance
are shown in Fig. 3b, for the case Z = Z$_{\odot}$. 
For Z = Z$_{typ}$, the results are 
similar, except for T $\leq$ 40000 K because of the effect of Fe on 
the gas cooling.
The curves in Fig. 3b show that 
the empirical N/H abundances are overestimated for T$_*$ $>$ 8$\times$10$^4$, 
while for lower T$_*$ the N abundance is lower than the true value. 
This behavior is related to the ionization correction factors used in 
the empirical calculations (see equations [1] to [4]).
In fact, for T$_*$ $<$ 8$\times$10$^4$ K 
the contribution of N$^{++}$ is underestimated since the correction 
for this non-observed ion is based on O$^{++}$ which has a smaller emitting
volume than N$^{++}$ (the volume of the O$^+$ 
zone being greater than the N$^+$ one due to charge transfer reactions).
The main effect for high T$_*$ 
is that the He$^{++}$ zone contains the N$^{++}$ zone; so, when applying the 
ionization correction factor for N$^{++}$, this ion is taken into account 
twice (both from the O$^{++}$ and He$^{++}$ corrections), increasing the 
obtained empirical N/H. 
Results for N/H from emission line intensities obtained with a slit 
crossing the nebular central region are greater than those where 
the whole nebula is observed. 
This comes from the fact that  with such a slit the observed emitting volume 
of high ionized zones relative to low ionized zones is larger than 
for the whole nebula. 

The behavior of the O/H empirical abundances with T$_*$ is similar to that of 
N/H, although less affected by the geometry. For example, when
compared to N, the O
abundance becomes greater than the true value for higher stellar 
temperatures. This leads to discrepancies of N/O which increase
with T$_*$ (Fig. 3b). 

Since He/H is one of the criteria for classifying Type I PNe, 
this ratio is shown in Fig. 4.
For He/H, empirical methods do not take into account H$^o$ nor He$^o$ 
abundances. However, for T$_*$ $<$ 4$\times$10$^4$ K, there is an 
increasing He$^o$ zone, larger than the H$^o$ one; when 
He$^o$ is not taken into account, lower abundances for He are then obtained. 
For T$_*$ $>$ 10$^5$ K, the H$^o$ zone is increasingly larger 
than the He$^o$ one due to the high energy photons which can ionize He$^o$ 
very far into the nebula. If this effect is not properly taken into account, 
larger He/H are obtained. 

For increasing values of n$_H$, the discrepancies of empirical values of N/H
increase, while those of He/H decrease (for increasing densities the volume 
of He$^o$ and H$^o$ tend to be equal). 
For example, for n$_H$ = 10$^4$ cm$^{-3}$, and same L$_*$ as in Fig. 3b, 
the empirical N/H ratio is higher by a factor more than 3 for T$_*$ $\geq$ 
2$\times$10$^5$ K.
Regarding the stellar luminosity, 
the effect is practically negligible for N, but the discrepancies of He/H 
increase for lower L$_*$.

\subsection{Bipolar nebulae}

The effect of the torus and of the resulting bipolarity on the [NII]/H$\alpha$
line ratio and on the corresponding empirical N/H are shown respectively 
in Fig. 5a and 5b. The curves correspond to T$_*$ = 10$^5$ K, 
L$_*$ = 3$\times$10$^3$ L$_{\odot}$, and Z = Z$_{typ}$.
Solid lines correspond to a spherically symmetrical nebula with a uniform
density (n$_H$ = 10$^2$ cm$^{-3}$) throughout the nebula. 
Dashed and dot-dashed curves show the result for a nebula with the same  
chemical abundance, T$_*$, L$_*$, and overall density, but including 
a torus around the central star. The torus density is n$_{torus}$ = 10n$_H$, 
and its dimensions are as described in \S 2. 
The dashed curve corresponds to a long slit along the lobes, and  
the dot-dashed one to a slit
perpendicular to the torus axis, both passing through the central star. 
For a bipolar 
nebula the [NII]/H$\alpha$ ratio as well as the resulting N/H vary with 
the position of the 
line of sight to the nebula, due to its geometry. However, the effect due
to the ionization distribution inside the nebula determined by the 
stellar temperature is much stronger, when the observations are made 
along the lobes, as it is often the case. 
For a given set of T$_*$, L$_*$, and overall density n$_H$, the 
main differences between the results for a bipolar and a 
homogeneous nebula are related to the torus characteristics, 
i.e, the contrast of n$_H$ with the torus density, as well as 
to the torus dimensions. For a slit along the maximum extent of the 
lobes, the main differences appear for lines of sight crossing 
the torus. The opening angle of the torus will be important for slits 
which make an angle with the torus axis.
Any intermediate slit position will show a variation between the 
dashed and dotted-dashed curves of Fig. 5a and 5b. 
The results presented in these figures are  
an illustration of the ambiguity on the determination of any 
abundance gradients using empirical methods on bipolar objects. 

\section{Conclusions}

The results of the preceding section show that 
empirical methods for abundance determination can simulate an 
overabundance and/or the presence of abundance gradients of He and N 
in planetary nebulae with high T$_*$, even when applied to
spherically symmetrical and homogeneous objects. This is also true for N/O. 
Recall that Type I PNe, 
defined as those nebulae which present an overabundance of He and N, have the 
hottest stars among PNe, and are generally bipolar.
As shown in \S 3, both characteristics can induce an apparent 
overabundance when empirical methods are used.

Following our results (Figs. 3a and 3b), systematic errors in the abundance
determinations by empirical methods are smaller for N/O. Thus, it is better 
to use the N/O ratio instead of N/H when discussing N rich planetaries. 
For central stellar temperatures in the range 10$^5$ to 3$\times$10$^5$ K the
N/O ratio can be overestimated by a factor of 1.5. On the other hand, the 
average N/O value for Type I PNe is about 2.8 higher than Type II PNe 
(Peimbert 1990). So, we expect that, even with a better N/O determination, 
Type I PNe may still be N rich compared to Type II PNe.

Inference about abundances on planetary nebulae are important
tests to stellar evolutionary models. 
Several arguments indicate that Type I PNe come from more massive progenitors
(Torres-Peimbert $\&$ Peimbert 1997). In particular, 
high masses for the central stars of Type I PNe are generally 
invoked to explain their high He and N abundances since the higher N/O 
ratio, the higher the original mass of the central star. 
Some theoretical evolutionary models fail to explain these overabundances
(Marigo et al. 1996). However, models accounting for the chemical evolution 
on AGB and including the "hot bottom burning" effect are consistent with 
PNe abundances (van den Hoek $\&$ Groenewegen 1997).
Our results show that He and N overabundances in Type I PNe may be lower than 
indicated by the empirical methods. The correction of the empirical 
abundances by those systematic errors change the constraints to the 
evolutionary models and may lead to different conclusions on the importance 
of the different processes. 

Furthermore, from the discussion in the 
preceding section, empirical methods applied
to PNe with high stellar temperatures can mimic abundance gradients.
Thus abundance gradients and their relation to multiple ejection phenomena
must also be reviewed. 

Let us recall that observed line intensities are integrated 
on the line of sight, and different 
lines of sight cross different fractions of regions differing
by their ionization state. Since lines originate in different regions, 
the observed image of a nebula depends on choice of the emission
line. As shown in the preceding section, abundances obtained from
observed line intensities depend on the line of sight.
Therefore, empirical abundances obtained from line intensities averaged
over different points in a nebula, or conclusions based on average 
abundances for a given nebula, may be incorrect.  

The effects shown in the present paper must be taken as a warning 
when discussing chemical enrichment and/or abundance 
gradients in a given nebula.
Any conclusion about abundances, in particular in nebulae with a high
star temperature, even with simple geometries,
must rely on detailed photoionization models which properly 
account for the integration of emission line intensities on a given 
line of sight. Classical unidimensional photoionization models 
are suitable for spherically symmetrical objects, whereas a 
3D photoionization code (Gruenwald at. al 1997) must be used for 
other geometries.

\acknowledgments
We are thankful to M. Peimbert for stimulating discussions and valuable 
suggestions. 
This work was partially supported by Fapesp, CNPq, and PRONEX/FINEP.

\newpage

\figcaption [] {Empirical abundance 
ratios from the literature (see text) versus the stellar temperature for 
N/H (a) and He/H (b).}

\figcaption [] {[NII]/H$\alpha$ line ratio (a) and the empirical N/H (b)
relative to the adopted value as a function of the 
line of sight.
The projected distance to the center of the nebula is given in units of the 
nebula ionized radius.
Curves correspond to simulated spherically symmetrical nebulae with L$_*$ = 
3000 L$_{\odot}$, n$_H$ = 10$^2$ cm$^{-3}$, and Z$_{typ}$, and are labeled 
by the stellar temperature in units of 10$^3$ K.}

\figcaption [] {(a) [NII]/H$\alpha$ versus T$_*$. 
The two upper 
curves correspond to Z$_{typ}$, while the lower ones to Z$_{\odot}$. 
Long-dashed and solid lines represent line ratios for the whole nebula, 
while dot-dashed and short-dashed curves correspond to line intensities 
integrated in a slit. The values of L$_*$ and n$_H$ are the same as in 
Fig. 2. (b) empirical N/H and N/O relative to the adopted value (Z$_{\odot}$).}

\figcaption [] {same as Fig. 3b, for He/H}

\figcaption [] { [NII]/H$\alpha$ (a) and N/H (b) relative to the adopted ratio
as a function of the line og sight, for T$_*$ = 10$^5$ K. 
Solid lines correspond to a spherically symmetrical nebula, while dashed and 
dot-dashed correspond to a bipolar nebula of same stellar temperature, 
respectively for a slit along the lobes, and a slit perpendicular to the 
torus axis.}

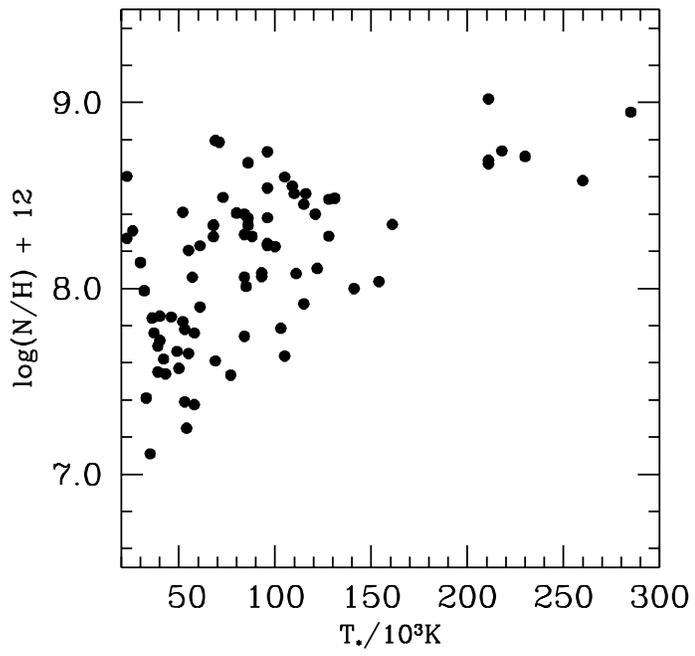
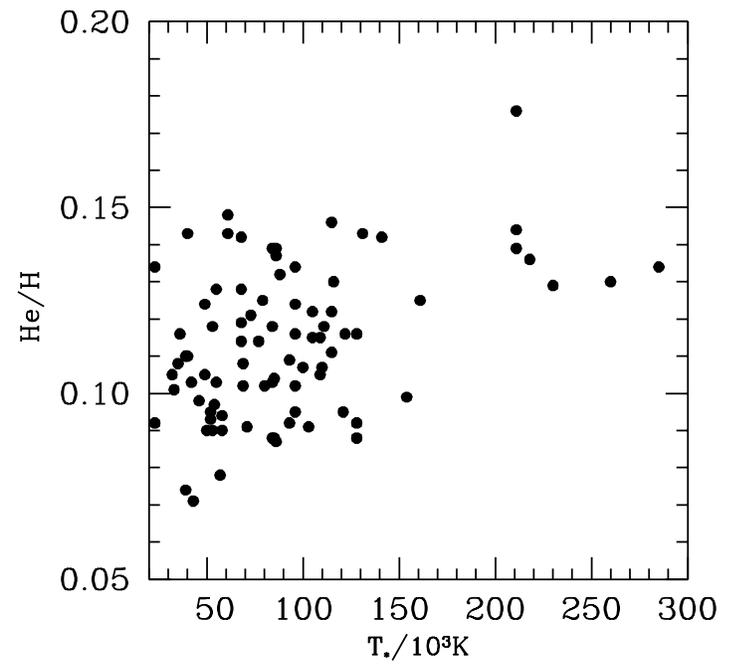
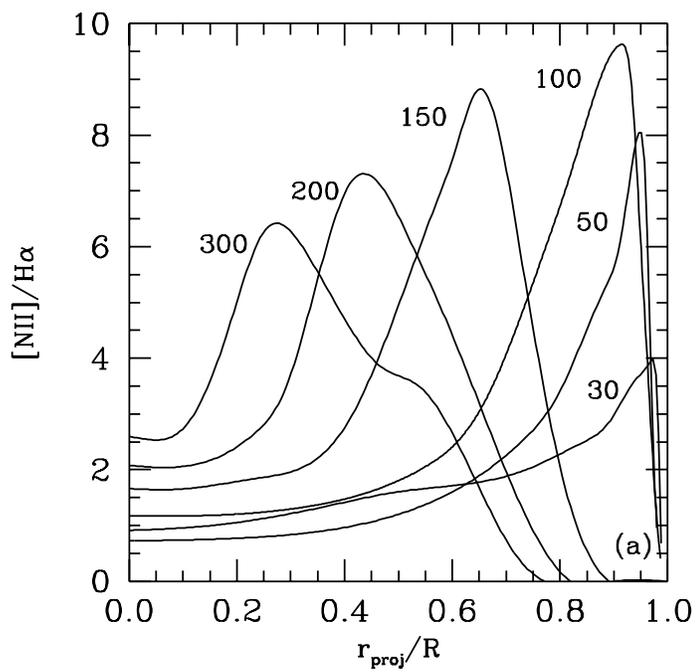
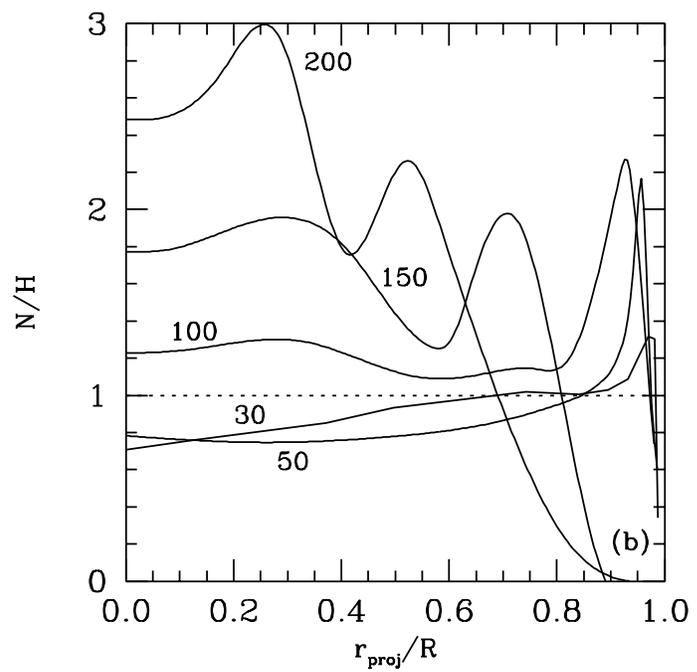
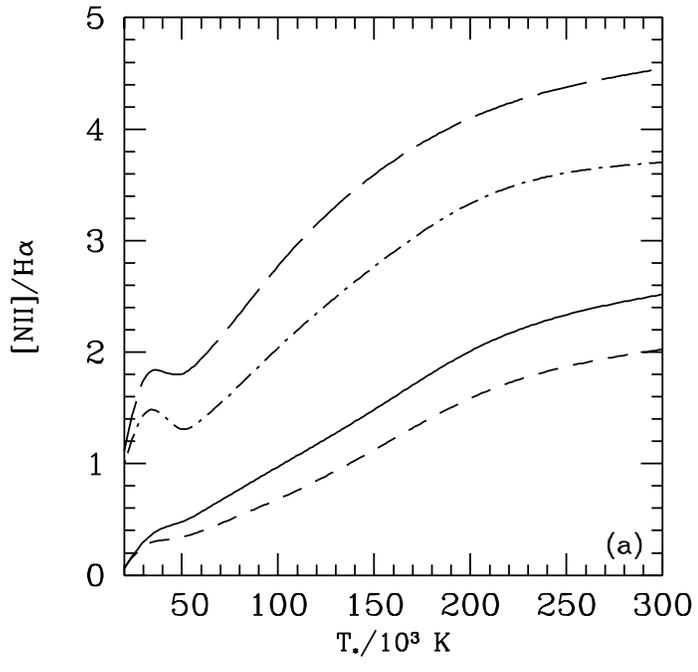
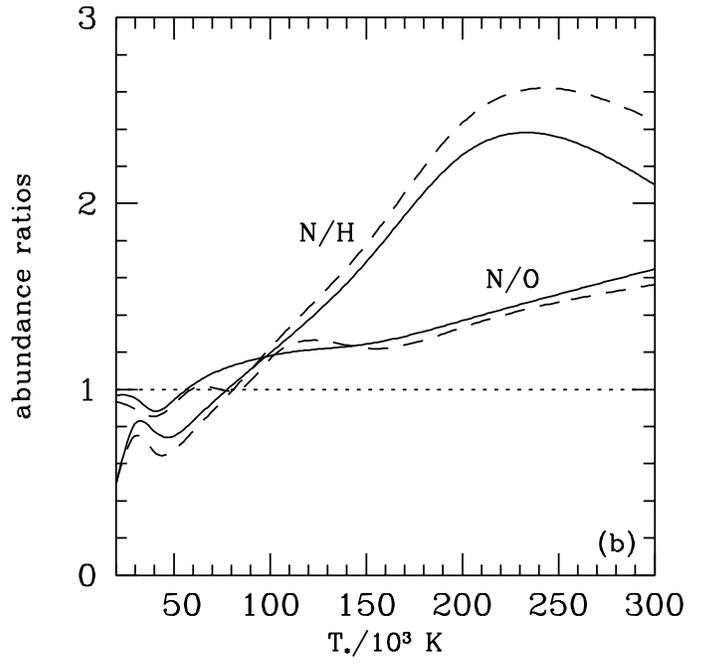
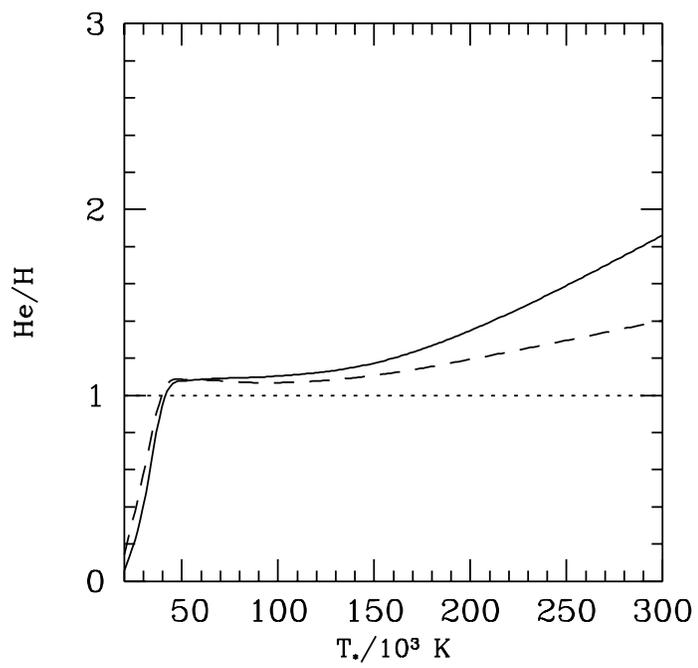
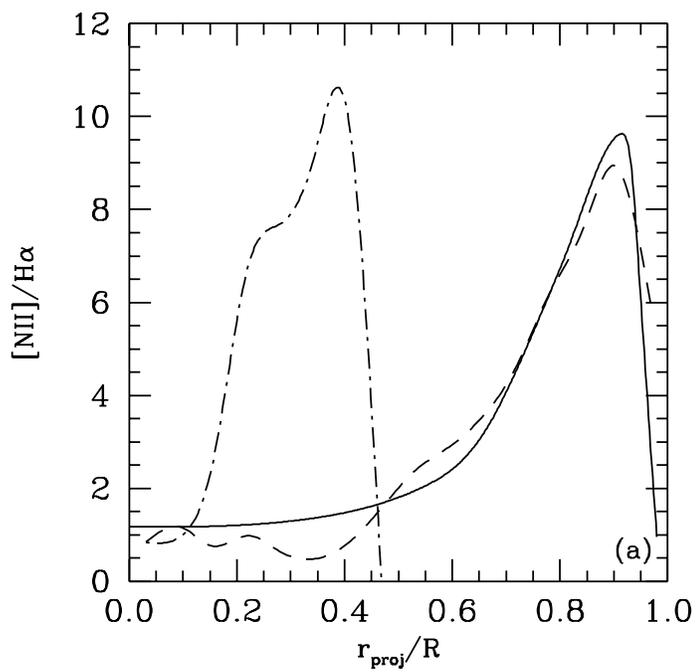
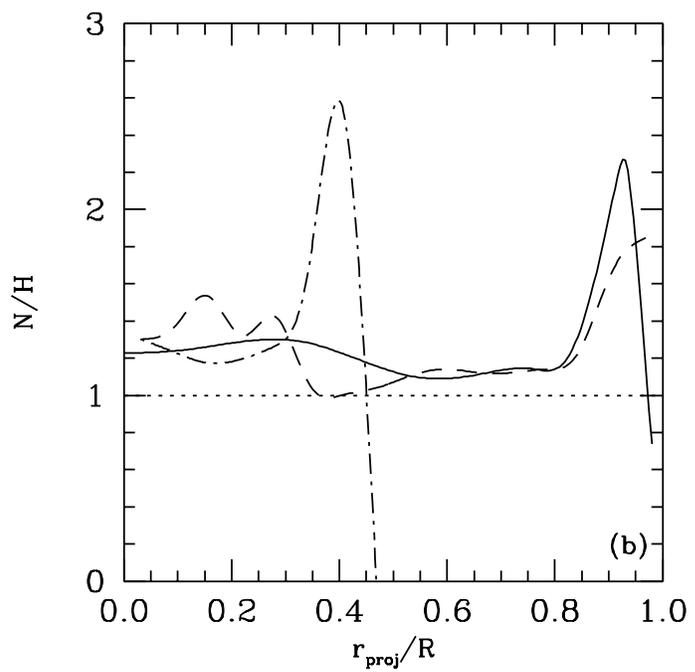
\end{document}